\documentstyle[epsf]{elsart}

\DeclareTextCommandDefault{\regtrade}{$^{\mbox{\scriptsize\textcircled{\tiny R}}}$}

\newif\iflindapicture

\begin{document}
\begin{frontmatter}
\rightline{FERMILAB-PUB-98/299-E}
\title{
%Construction, Operation, and Performance of 
The  SELEX Phototube RICH Detector}
\author[fermi]{J.~Engelfried\thanksref{slp}},
\author[msu]{I.~Filimonov\thanksref{dec}},
\author[fermi]{J.~Kilmer}, 
\author[ihep]{A.~Kozhevnikov}, 
\author[ihep]{V.~Kubarovsky}, 
\author[ihep]{V.~Molchanov},
\author[msu]{A.~Nemitkin}, 
\author[fermi]{E.~Ramberg}, 
\author[msu]{V.~Rud},
\author[fermi]{L.~Stutte}
\address[fermi]{Fermi National Accelerator Laboratory, 
Batavia, IL, USA\thanksref{DOE}}
\address[ihep]{Institute for High Energy Physics, Serpukhov, Russia\thanksref{RMS}}
\address[msu]{Moscow State University, Moscow, Russia\thanksref{RMS}}
\thanks[slp]{Now at Instituto de F\'{\i}sica, 
Universidad Autonoma de San Luis Potos\'{\i}, Mexico}
\thanks[dec]{deceased}
\thanks[DOE]{Work supported by the US Department of Energy under
contract NO.\ DE-AC02-76CHO3000.}
\thanks[RMS]{Supported by the Russian Ministry of Science
and Technology.}
\begin{abstract}
In this article, construction, operation, and performance of the RICH
detector of Fermilab experiment~781 (SELEX) are described.
The detector utilizes a matrix of
2848~phototubes for the photocathode
to detect Cherenkov photons generated in a $10\,\mbox{m}$ Neon radiator.
For the central region an $N_0$ of $104\,\mbox{cm}^{-1}$,
corresponding to $13.6\,\mbox{hits}$
on a $\beta=1$ ring, was obtained.  The ring radius resolution measured
is $1.6\,\mbox{\%}$. 
\end{abstract}
\end{frontmatter}

\section{Introduction}
The Fermilab experiment E781 (SELEX): 
A Segmented Large $x_F$ Baryon Spectrometer~\cite{e781,charm2000},
which took 
data in the 1996/97 fixed target run at Fermilab, is designed to
perform high statistics studies of production mechanisms and decay physics of
charmed baryons such as
$\Sigma_c$, $\Xi_c$, $\Omega_c$ and $\Lambda_c$\@.
The physics goals of the experiment require good charged
particle identification to look for the different baryon decay modes.
One must be able to separate $\pi$, $K$ and $p$
over a wide momentum range when looking for charmed baryon decays like
$\Lambda_c^+ \rightarrow p \, K^- \pi^+$.

A RICH~\cite{ArtR} detector with a 2848~phototube photocathode array
has been constructed~\cite{protos,sphinx} to do this. The detector begins
about $16\,\mbox{m}$ downstream of the charm production target, with two
analysis magnets with $800\,\mbox{MeV/c}$ $p_t$-kick each in-between, and is
surrounded by multi-wire proportional and drift chambers which provide particle
tracking.  The average number of tracks reaching the RICH is about 5 per event.
First results from this detector can be found in~\cite{elba}.

In this article we first describe the properties of the main 
parts of the detector
(vessel, mirrors, photon detector). Finally we report about 
stability during the run, performance, and some preliminary physics results.

\section{Vessel}
\begin{figure}[htb]
\begin{center}
\leavevmode
\epsfxsize=\hsize
\epsfbox{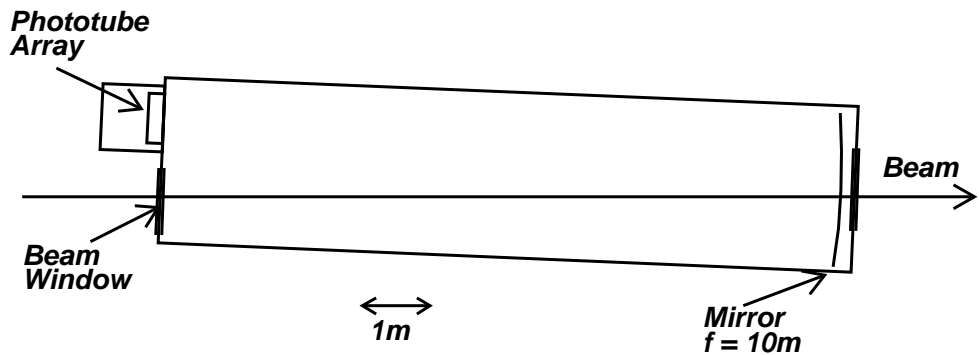}
\caption{Vessel Layout.}
\label{vessel}
\end{center}
\end{figure}

The E781 RICH vessel is a low carbon steel vessel $10.22\,\mbox{m}$ in length,
$93\,\mbox{in.\ }$in diameter and with a wall thickness of 
${1\over{2}}\,\mbox{in.}$
The end flanges are 
$1.5\,\mbox{in.\ }$thick aluminum with provisions for thin beam 
windows and a
phototube holder plate to be described later.
The heavy construction\footnote{Rode Welding 
Service Inc., Elk Grove Village, IL.} allowed the vessel to be
leak checked with a helium mass spectrometer.  The vessel specification
was to show no leak greater than $10^{-9}\,\mbox{atm\,cc/s}$. 
It easily passed the
leak test and proved to be a very tight vessel in operation.

\looseness=1
The vessel has three feedthroughs and several piping connections for the
gas system.  Electrical feedthroughs are gas-tight vacuum coaxial
connectors.  Provisions were made for thermistors for both ends of the
vessel to monitor internal temperature and an LED pulser test array to
supply calibration signals for the phototubes.  The third port has
connections for the gas analyzers.

The entire interior of the vessel was painted with a flat black paint to
reduce reflections of photons in the vessel.  To keep the paint from
contaminating the radiator gas a very low out-gassing 
paint\footnote{Lord Corporation
Industrial Coatings, Aeroglaze~Z306
polyurethane flat black paint.}
was chosen~\cite{nasa}. 
The inner surface of the vessel was
sand-blasted before painting and then sprayed.

Several other mechanical features of the vessel are significant.  The
vessel is tilted off the horizontal axis by
$2.4\,\mbox{deg}$ with the beam 
entering horizontally through a thin window.  The RICH is also mounted on
a rail system that allows it to be moved out of the beam line for the alignment
of the mirrors and for certain
types of physics running.
Stops on the rails were adjusted to allow
the vessel to be put back on the beam line repeatably within $\pm1\,\mbox{mm}$
after a vessel move.

\begin{sloppypar}
The entrance and exit windows are 
fabricated\footnote{Shelldall, Northfield, MN.}\ from a 3-layer laminate
that consists of $3\,\mbox{mil}$ Kevlar\regtrade\ cloth, 
$1\,\mbox{mil}$ aluminum 
foil and $1.5\,\mbox{mil}$
charcoal Tedlar\regtrade.  
These three elements combined give a strong, gas-tight 
and light-tight barrier.  Permeability tests were conducted with
both air and neon gas samples, and showed results consistent with the
background level of the measuring apparatus.  Light-tightness was
checked by measuring the noise rate in a phototube placed in a box with
one side made of the window fabric, and showed no observable increase
from that of a phototube inside a light-tight box. For mounting on the vessel,
the window material was sandwiched between two mylar rings, and inserted
between two aluminum rings which were bolted together along with a rubber
o-ring seal.  
\end{sloppypar}

To avoid the construction of a complicated gas-filling 
system to prevent the collapse
of the phototube holder plate,
which can only support 
a pressure difference across it of $3\,\mbox{psid}$,
a new gas system was 
constructed
to purge the vessel which requires only an operating
pressure of $1\,\mbox{psig}$~\cite{regenia}.
The vessel was purged by flow purging
from air to carbon dioxide.  Carbon dioxide was introduced at the bottom
of the vessel and displaced air was removed from the top.   After the air
was removed a separate circuit of the system added 
neon to the now $\mbox{CO}_2$ filled vessel and 
froze out the excess $\mbox{CO}_2$ with a liquid nitrogen cooled
freezer keeping the pressure constant.  When the vessel contained neon
with only a trace of $\mbox{CO}_2$ 
the final step was to pump the vessel gas through a
liquid nitrogen cooled adsorber to remove the last traces of
contaminates.  In the initial purge for the run the oxygen background was
reduced to $3\,\mbox{ppm}$.  
The oxygen level was data logged throughout the run.  Over
the course of a running period lasting more than a year the oxygen level
rose to an estimated $(20\pm 12)\,\mbox{ppm}$.  The large error in the final
$\mbox{O}_2$ level stems from the fact that the 
oxygen monitor failed in the
middle of the run due to aging; replacing the cell after the run introduced
some oxygen into the system and resulted in a reading of 
$32\,\mbox{ppm O}_2$\@. 
An extrapolation from the last accurate reading was $8\,\mbox{ppm}$.

The gas system was designed to allow the removal of oxygen from the
neon gas during a running period. Because of the tightness of the vessel
and gas system the oxygen contamination never became severe enough to
require purifying the neon.  As a result the initial gas charge remained
without change in the vessel for the entire run.

The vessel also has a
constant temperature control system provided by a water coil around the
vessel wall and a chilled water system.  
Fifteen cm 
thick building insulation 
was used
to help keep the vessel temperature
constant.  During the running period the temperature was set to 
the same temperature as when the mirror mounts were glued onto the mirrors and
the mirrors were
installed and aligned.  Over the course of the run (15~months) the
temperature was  $21.5\,^\circ\mbox{C}\pm 1.5\,^\circ\mbox{C}$.  

To monitor the status of the detector, data was logged~\cite{epicure}
for the entire run
once per minute.
Parameters monitored were:
atmospheric pressure, vessel pressure, vessel temperature,
oxygen content, temperature at several locations within the phototube box,
temperature in the digitizer crates,
voltages at the end of the HV Zener diode chains,
and status of the low voltage system.  Error conditions were automatically
checked and reported to the shift crew.

\section{Mirrors}
\begin{figure}[htb]
\begin{center}
\leavevmode
\epsfysize=10cm
\epsfbox{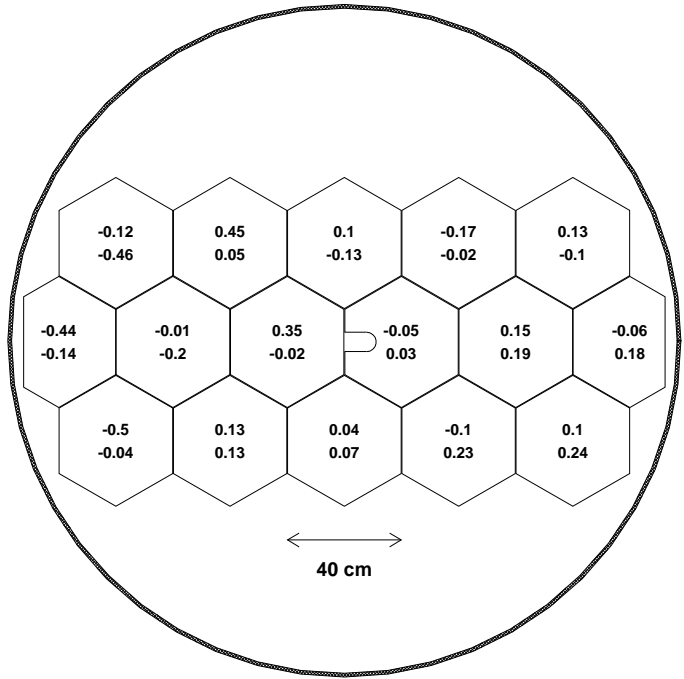}
\caption{Mirror Layout in the Vessel.  The numbers show the deviation from
the average center of curvature for the horizontal (top) and vertical
(bottom) coordinate in $\mbox{cm}$.}
\label{mirrorhex}
\end{center}
\end{figure}

The mirror plane at the end of the vessel is build up of 16~hexagonally 
shaped spherical mirrors, covering a 
total area of $2.4\,\mbox{m}\times 1.2\,\mbox{m}$, as shown in 
Fig.~\ref{mirrorhex}.
Each mirror is $40\,\mbox{cm}$ 
across ($46\,\mbox{cm}$ tip-to-tip), has a thickness of $10\,\mbox{mm}$,
and is made of low expansion glass\footnote{Schott Tempax Glass, 
provided by Abrisa Industrial Glass Inc., Venture, CA.}.
The average radius of the mirrors is $1980\,\mbox{cm}$, with a deviation
of $<5\,\mbox{cm}$~RMS between mirrors as well as on one mirror.  The mirrors
were polished\footnote{Astronomically Xenogenic Enterprises (AXE), Tucson, AR.}
until the image of a point source illuminating the
whole mirror was $<1\,\mbox{mm}$. The back of the mirror is rough polished to
the same radius.

%\looseness=1
The leftmost and the rightmost mirrors in the central row had to be
truncated by
$3\,\mbox{in.\ }$to fit into the vessel.  The central mirror has a cutout
of $6.8\,\mbox{cm}\times 11\,\mbox{cm}$ to allow non-interacting beam particles
to pass through the detector without interacting in the mirror to reduce bad
effects in more downstream detectors.

The mirrors were coated\footnote{Acton Research Corporation, Acton, MA.} with 
Aluminum, and
an 
overcoating of $\mbox{MgF}_2$.  The reflectivity was
$>85\,\mbox{\%}$ at $160\,\mbox{nm}$.
 
The sphericity of all mirrors was measured using the 
Ronchi method~\cite{Ronchi}, which confirmed that the mirrors were
produced within the specifications.  The results were used to mount the
mirrors with the best sphericity in the center, and to have a minimal 
difference of the average radius between neighboring mirrors.

\begin{figure}[htb]
\begin{center}
\leavevmode
\vspace{11cm}
\includegraphics{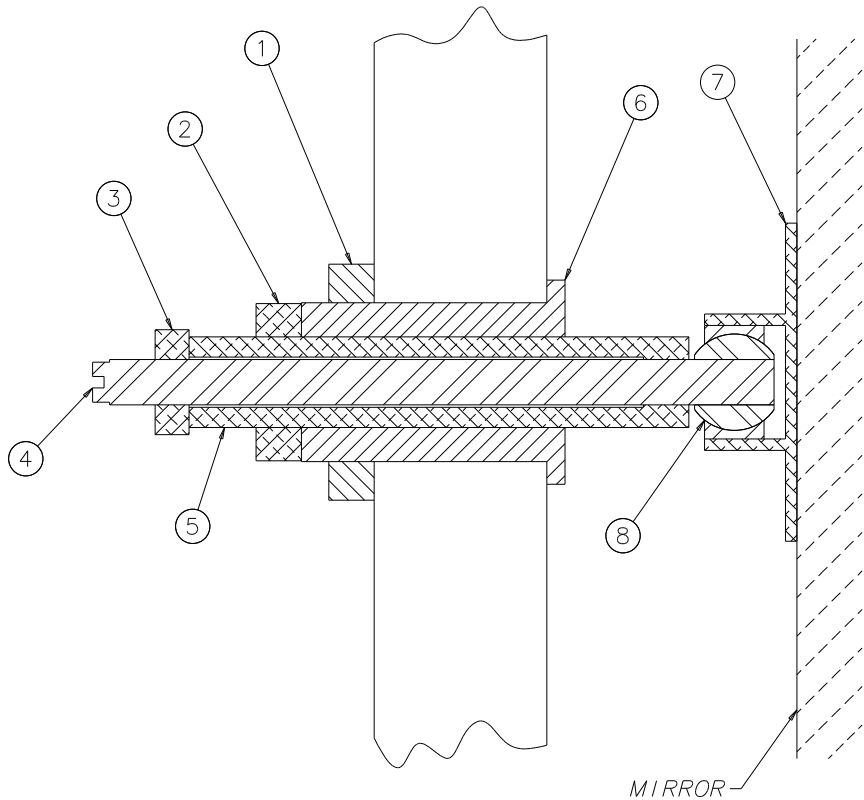}
\caption{Mirror Mount. The honeycomb panel is $1\,\mbox{in.\ }$thick.
1: Nylon nut. 
2,3: Aluminum nuts.
4: Titanium rod.
5: Aluminum rod.
6: Nylon cylinder.
7: Aluminum pad glued to mirror.
8: Ball bearing.
}
\label{mirrormount}
\end{center}
\end{figure}

\begin{sloppypar}
The mirrors are fixed individually to a flat, low mass 
honeycomb panel\footnote{Plascore Inc., Zeeland, MI.} of
$1\,\mbox{in.\ }$thickness with a 3~point kinematic
mount~(see Fig.~\ref{mirrormount}). 
Each mount consists of an aluminum pad 
glued\footnote{BIPAX\regtrade TRA-BOND BA-2151, TRA-CON, Medford, MA.} 
to the back of the
mirror, a ball bearing, and a double differential screw built out of
a $\quart\,\mbox{in.\ }$diameter titanium rod with a thread of 
$80/\mbox{in.}$, a $\half\,\mbox{in.\ }$aluminum 
cylinder with a thread of $13/\mbox{in.\ }$and a nylon 
cylinder of $1\,\mbox{in.\ }$diameter.  This mounting scheme allowed    
alignment of the mirrors on a sphere with a maximum sagitta 
of $2\,\mbox{cm}$, while
still maintaining an angular accuracy of $50\,{\mathrm \mu rad}$. 
\end{sloppypar}

Due to space restrictions in the experimental area, the mirrors had to be
mounted and aligned with the vessel moved laterally by $3\,\mbox{m}$ 
from its final location in the experiment.  Great care was taken in assuring
that vibrations during movement of the vessel do not misalign the mirrors.
After mounting the mirrors on the honeycomb panel, a first rough alignment 
was done by eye.  The final alignment was performed with a 
laser mounted on a theodolite base, sited at the average center of
curvature.  The mirror angles were adjusted until the reflected spot
observed back at the center of curvature had no more 
than $2\,\mbox{mm}$ displacement.

To determine the alignment of the mirrors during the run, standard
experiment data (no special data sets are needed) 
were used to measure the center 
of curvature of each mirror.
In an iterative procedure
the predicted ring centers (using the previous
set of constants) are compared to the centers 
obtained from a circle fit to isolated
rings, using only tracks where the Cherenkov light was reflected by one
mirror. This procedure converges after typically 5 iterations. With
a typical data set of 1~million events, between 2000 and 50000 tracks are used
to align the different mirrors.  This alignment procedure was repeated
for data sets taken over the whole running time of the experiment and it was
found that the obtained results are constant.
The final deviations from an average center of curvature are shown in 
Fig.~\ref{mirrorhex}.  The mean lateral displacement is $2.6\,\mbox{mm}$.

\section{Photodetector}
\subsection{Holder plate}
\begin{figure}[p]
\begin{center}
\leavevmode
\epsfxsize=\hsize
\epsfbox{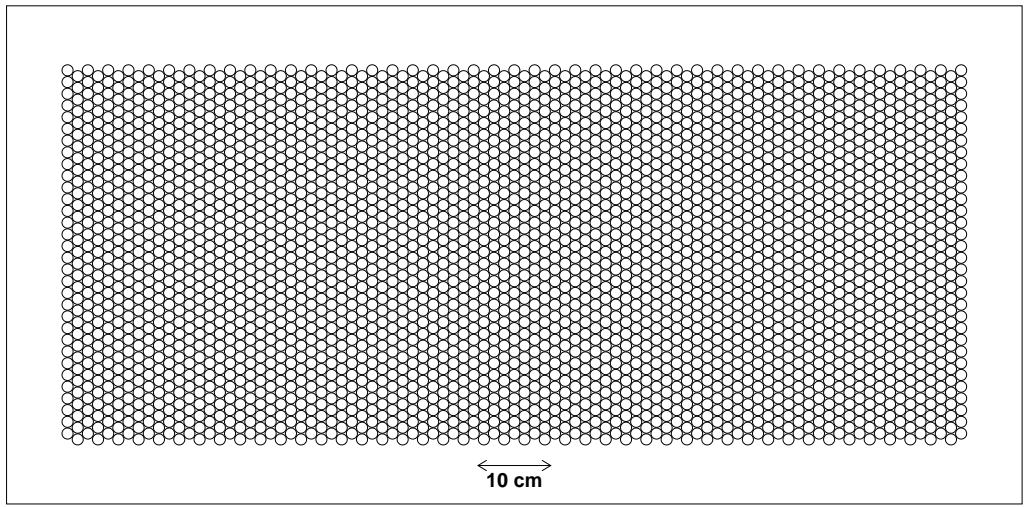}
\vskip 1cm
\epsfxsize=\hsize
\epsfbox{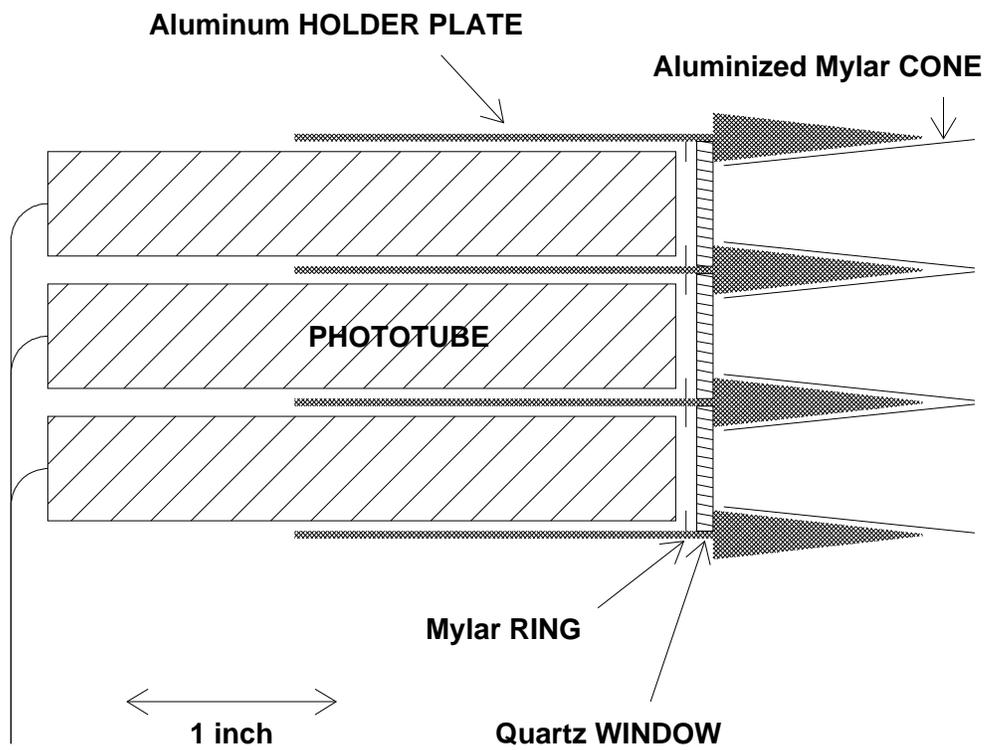}
\caption{Front view (top) and partial cross section (bottom) through 
the Phototube Holder Plate.}
\label{holderplate}
\end{center}
\end{figure}

\begin{sloppypar}
The phototube holder plate is 
made\footnote{Walco Tool and Engineering Corp., Lockport, IL.} 
from a $55\,\mbox{in.\ }$wide,
$27\,\mbox{in.\ }$high, $3\,\mbox{in.\ }$thick aluminum block. 
Figure~\ref{holderplate} 
shows a cross section of a few of the
 $2848$ ($89\times 32$) hexagonally packed ($0.635\,\mbox{in.\ }$spacing)
holes of 
approximately $0.6\,\mbox{in.\ }$diameter that were drilled through it.
\end{sloppypar}

\looseness=1
One side of each
hole is a straight channel $2\,\mbox{in.\ }$in depth, 
which is used to support
a phototube.  A quartz window\footnote{Suprasil~2 from Heraeus Amersil Inc.,
Duluth, GA.
}
of $2\,\mbox{mm}$ thickness was inserted into 
this side of the hole and glued in place, providing a gas seal between the
neon radiator and the phototube housing.  
After the windows were glued in place, they were checked for
gas-tightness using a helium leak detector.  A small number of 
leaky ($> 10^{-8}\,\mbox{atm cc/s}$)
windows were replaced.  All windows were then cleaned of residual glue
(which was not UV transparent) and their transmission checked using a
spectrophotometer equipped with a fiber-optic system interface.

\looseness=1
The other side of each hole is
a tapered channel $1\,\mbox{in.\ }$in depth, 
with an inner radius of $0.4\,\mbox{in.}$
An aluminized mylar cone is inserted into this side and extends slightly
out from the block in order to give essentially $100\,\mbox{\%}$
coverage for photon detection. 
The $0.4\,\mbox{in.\ }$inner 
radius of the cone section of the hole was made to be the
same size as the photocathode of the phototubes.  In order to center the
tubes on the aperture and provide enough friction to hold them snugly in
place, thin Velcro\regtrade\ strips (loop half) were affixed along each tube.  
To minimize the loss of UV photons, the gap between the quartz window and
the phototube had to be kept as small as possible.  The FEU60 phototubes
however were coated with PTP wavelength shifter, so a
small gap was needed in order to avoid
removing the coating.  These two goals were achieved by placing a thin
($3\,\mbox{mil}$) mylar ring of 
inner diameter $0.4\,\mbox{in.\ }$up against the window. 

\subsection{Phototubes}

Two different types of ${1\over{2}}\,\mbox{in.\ }$diameter, 
$10\,\mbox{mm}$ bi-alkali photocathode,
10-stage
tubes are used.  The first is a commercially available tube 
R760\footnote{Hamamatsu Corp., Bridgewater, NJ.}
which has a quartz window and thus response down to 
$170\,\mbox{nm}$.  It has
a quantum efficiency of approximately $25\,\mbox{\%}$
at its peak wavelength of $350\,\mbox{nm}$.
The second is a Russian tube (FEU60) which has a glass entrance window.
These tubes were coated with PTP wavelength shifter to reach the same 
wavelength range as the R760 tubes (the quartz entrance window of the
holder plate provided the 
$170\,\mbox{nm}$ wavelength cutoff).
Initial
tests in a vacuum reflectometer equipped with a Deuterium VUV-UV light
source showed that the thickness of the PTP was
not a critical parameter for their performance, so as many as 70~tubes
were coated at one time in a vacuum deposition jar.

Each of the phototubes in the SELEX RICH differed
with respect to their performance under high voltage.
Because each column of 32~phototubes was connected to a single
high voltage, it was thus required to determine each phototube's
characteristics and match it to 31 other similar tubes.

The two types of tubes, R760 and FEU60, are dramatically
different in their quantum efficiency (Q.E.) and noise characteristics.
The FEU60 tubes are on average only $42\,\mbox{\%}$ as efficient 
and have at least an order of magnitude greater noise at 
operating voltage than the R760 tubes. 
Because it was desired to limit the accidental
noise in the detector to less than $0.5\,\mbox{\%}$,
a simple algorithm was developed to determine operating
voltage:  either the voltage where the Q.E.\ 
plateaued, or the voltage where the tube's inherent noise reached
$30\,\mbox{kHz}$.  This would limit noise in the detector up to our
specification for gate values up to $170\,\mbox{ns}$.

\looseness=+1
The operating parameters of the phototubes were determined with a 
test setup where 
32~phototubes could be measured at once.
Quantum efficiencies were
measured relative to a standard R760 phototube, whose absolute
Q.E.\ was only approximately known.
The setup used a blue LED, whose light
was delivered to 32~identical phototube holders and the standard
tube via optical fibers,
all inside a dark box.  The LED was driven such that it delivered
on the order of $5\,\mbox{\%}$ of a photoelectron per pulse to each phototube.
The output of the phototubes were digitized with the 
hybrid chips used in the detector,
and the ECL output coming from
this readout was sent to two CAMAC scalers in series.  One scaler was
gated for $100\,\mbox{ns}$ 
in time with the signals corresponding to the LED light.  The
other scaler was gated for the same amount of time, but
midway between successive LED pulses, to monitor
the inherent noise of the phototube.  This noise level was subtracted
from the in-time scaler count to provide the true in-time response of
the phototube.  Another scaler kept count of the number of LED pulses
to provide a normalization.  The phototube
voltage for all 32~test tubes was provided by a 
CAMAC-controlled high voltage supply\footnote{LeCroy Corp., 
Chestnut Ridge, NY\@. Model 4032}
and was scanned over the
operating voltage range
for all tubes.  
Approximately $5\cdot10^5$ LED pulses were taken for each data point.

\looseness=+1
After testing all phototubes, a computerized matching program was
developed to associate 34~phototubes having the same characteristics
into a single batch with one operating point for high voltage.  Two
phototubes were kept as spares and the remaining 32~were bundled 
together in two batches for installation in the detector.
In the central part of the phototube holder plate 
the R760 and the FEU60 tubes were installed in alternating columns; 19 columns
are equipped with R760 tubes. In the outer parts,
only FEU60 tubes were installed.

\subsection{High Voltage, Low Voltage, and Readout Systems}
\begin{figure}[htb]
\begin{center}
\leavevmode
\epsfxsize=\hsize
\epsfbox{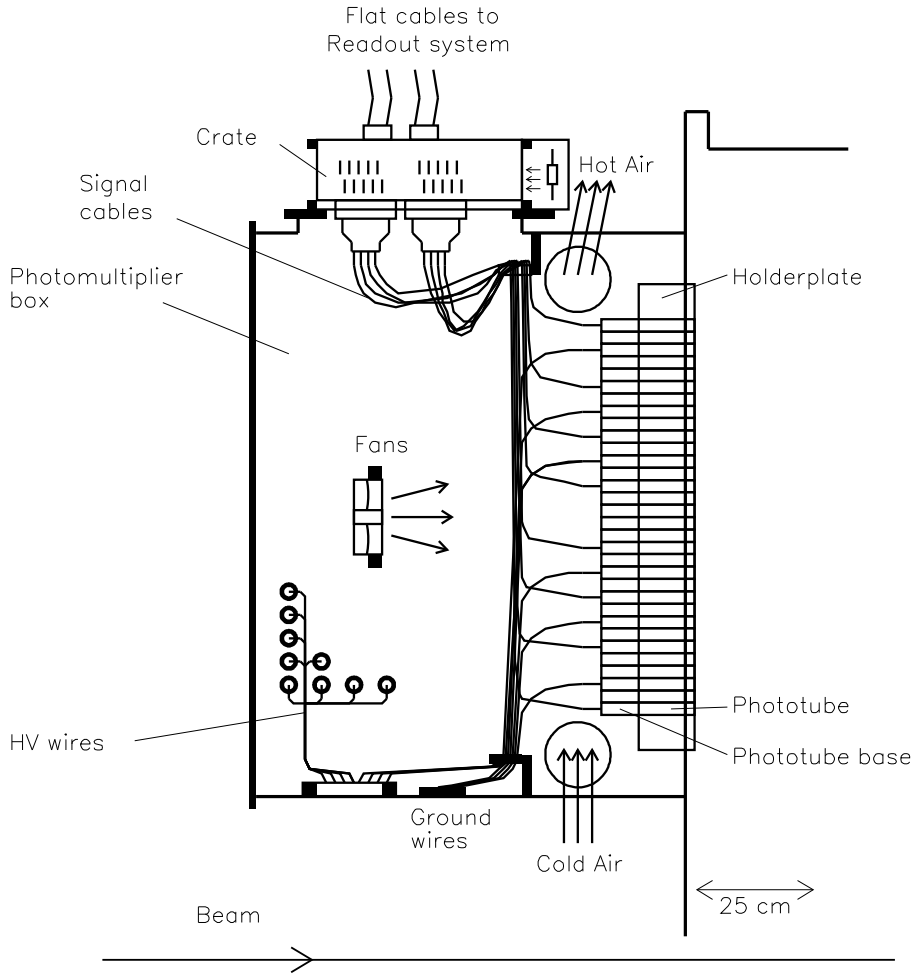}
\caption{Phototube Box.}
\label{elecbox}
\end{center}
\end{figure}

\looseness=+1
To accommodate the wide range of operating voltages for the phototubes 
(FEU60: $1300\,\mbox{V}$ to $1900\,\mbox{V}$ with a current of
$\approx 150\,{\mathrm \mu}\mbox{A}$ per tube, 
R760: $900\,\mbox{V}$ to $1250\,\mbox{V}$ with a current draw of 
$\approx 300\,{\mathrm \mu}\mbox{A}$ per tube) and the large number of
tubes, six chains of air-cooled Zener diodes were used,
each driven by a high voltage
power supply\footnote{Manufactured in Russia.} delivering $100\,\mbox{mA}$.
The voltage for one of the 89~columns is selected at the proper location
within the Zener diode chains. This distribution system is located outside
the phototube box (Fig.~\ref{elecbox}). 
Inside the box, the cables are soldered together with
the wires to each of the 32~phototubes on a small fanout board. The ground
returns are collected as well via fanout boards and are connected in common
to a feedthrough panel.  A support system holds the fanout boards in place.

\looseness=+1
An interlock system is used so that the high voltage is shut off
   before the cover can be removed from the phototube box.  Four separate
   rocker switches (one at each corner of the box) will detect whether or
   not the box is closed.  In addition, signals from
several temperature sensors located inside the box to detect 
any unusual conditions contribute to the interlock.

About $1\,\mbox{kW}$ of heat is generated inside the box by the phototubes.
   A simple cooling system was designed to remove the heat using chilled
   water with a light-tight heat exchanger.  The cooled air enters through a 
   long slit in the bottom of the box near the phototubes. There are two 
   returns for the air, each located near the top sides of the box. Fans
are located inside the box to improve the airflow.

The phototubes are grouped in sets of 16, two sets comprising a column
of 32~phototubes in the $89\times 32$~matrix.  The output signals from a group
of 16~phototubes are soldered onto paddle cards.  These cards
contain a $1\,\mbox{k}\Omega$~resistor 
for each channel which protects the corresponding
readout chip from charge build-up whenever the paddle cards are initially 
connected.  The paddle cards plug into a lighttight backplane of 
one of three custom made crates, 
located on top of the phototube box (Fig.~\ref{elecbox}).
96-pin Eurocard connectors\footnote{AMP Inc., Harrisburg, PA\@. Type 1-215913-4.}
are used in the backplanes. To prevent
cross talk between the signal wires, only every third column in the connector
is used, alternating the signal and ground pins;  all unused pins are grounded.

The readout electronics, mounted on cards in the crates, consists of hybrid 
chips\footnote{Designed by Moscow State University. Manufactured in
Russia and Hybrids International, Ltd, Olathe, KS.}
containing a 
preamplifier, a discriminator and an ECL line driver. 
The $\pm 6\,\mbox{V}$ operating voltages for the chips are
provided by six power 
supplies\footnote{Lambda Electronics Inc., Melville, NY\@.
LXS-7-6-OV ($60\,\mbox{A}$).}
and distributed to the cards 
via the backplane 
with highcurrent connectors\footnote{AMP Inc., Harrisburg, PA\@. 
Types 207541-3 and 207609-3.}.
Fans are mounted on each crate for cooling the hybrid chips.  
Air flow sensors are located in the air stream for each crate to monitor 
this cooling and to provide a safety switch-off.

During the initial phase of installation of the hybrid chip readout,
the cooling system was not adequate to keep the chips at a modest
temperature.  During 
the run 
many chips stopped working.  An analysis of these chips indicated
equal failures of the amplifier and discriminator.  Although the cooling
system was improved, chips
continued to fail at the rate of
several per week.
Spare chips 
were used to replace failed chips on the cards.  However,
the population of these spare chips 
was not adequate to deal with the chip failure rate. An
alternative readout card was developed 
that could be plugged into the same electronics
infrastructure and that delivered the same type of output.  This card 
consisted of a preamplifier circuit board with discrete components, leading
into a Nanometrics N277\footnote{Nanometric Systems Inc., Oak Park, IL.} 
discriminator and ECL driver card.
These
cards replaced 32~channels at a time and freed up hybrid chips.
At the end of the run, about 1/4 of the channels were serviced by N277 cards.

The $40\,\mbox{ns}$ wide digital outputs 
from the hybrid chips and the N277 cards 
are routed to ECL Latches~\cite{CROS}, which were actually
$1\,\mbox{bit}$ shift registers for every channel.  The shifting is performed
every $57\,\mbox{ns}$ ($1\over 3$ of the Tevatron radio frequency).  3~slices
were combined for a total integration time of $170\,\mbox{ns}$.
No attempt was made to shorten the integration time, since the detector
showed a very low noise rate. 
The zero suppressed readout yielded only the addresses of hit channels.

\subsection{LED Monitoring System}
 The LED monitoring system was implemented for off-spill monitoring and checks 
of the photocathode and readout. This system illuminates the photocathode 
surface with short light pulses from blue LEDs mounted onto the mirror support 
frame. It also includes LED driver circuits (to ensure short light pulses),
separate for each LED, fed from a $12\,\mbox{VDC}$
power supply, and a pulse generator.

The LED drivers are connected in four chains, about 10~LEDs in each. Only two 
chains are used, the others are spare. The whole chain is fired simultaneously 
from a NIM pulse generator either manually, or by computer control. 
The amount of light produced by the LEDs can be controlled via the 
pulse amplitude manually. The distance between LEDs and photocathode is rather 
big (the vessel length), giving a very even distribution of light on the 
photocathode surface. Short pulses give about the same phototube response 
as Cherenkov light, so the monitoring conditions are 
very close to the working ones. 
 
 An interspill monitoring task ran automatically during the whole run
and produced a list 
of "hot" and not working channels. To  determine noise, a readout was
initiated 
without firing the LEDs. 
Dead and hot channels were stored in a database and are used in the 
offline analysis to adjust parameters used in the likelihood function (see
later).

\section{Determination of Performance Parameters}
In this section we define and determine the operation and performance 
parameters of the detector.  After a description of the particle
identification algorithm, we determine the Figure of Merit $N_0$, the
refractive index, the single hit resolution $\sigma_h$, and the 
ring radius resolution $\sigma_r$, as a function of time (where needed).

\subsection{Particle Identification Method}
\begin{figure}[htb]
\begin{center}
\iflindapicture
\leavevmode
\epsfxsize=\hsize
\epsfbox{ricpix0183.eps}
\else
\vspace{10cm}
\includegraphics{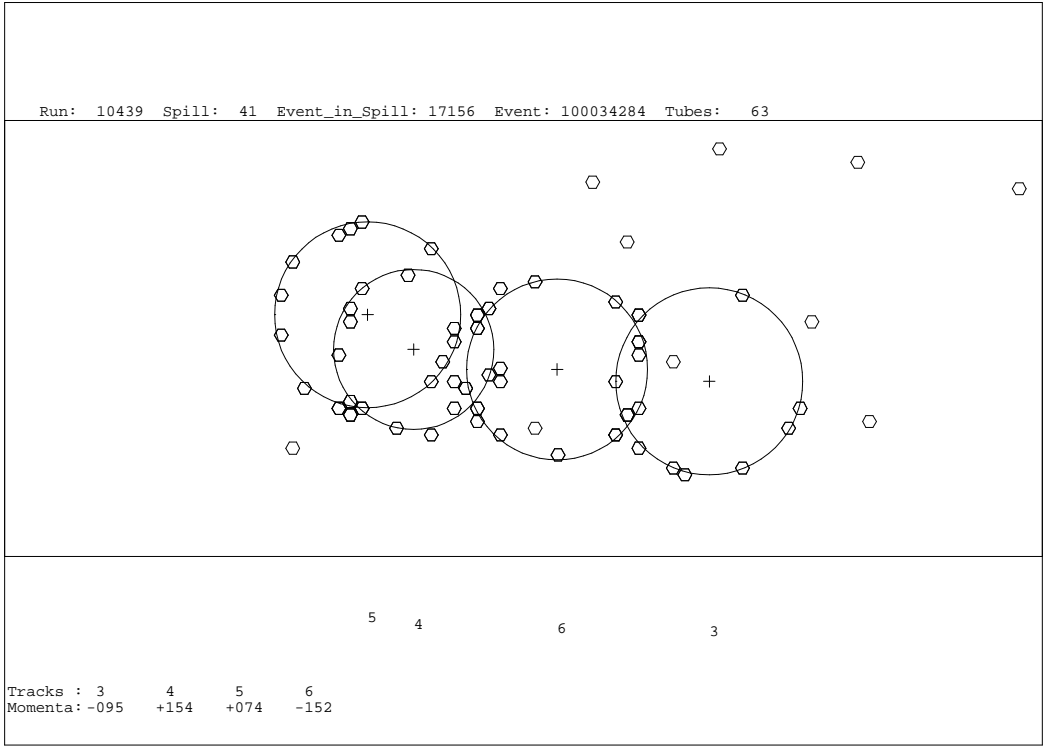}
\fi
\caption{Single Event Display. The small hexagons represent a hit phototube,
the circle show the ring for the most probable hypothesis, the numbers below
the rings denote the track numbers.
\iflindapicture
The vertical lines show the division of the photocathode into different
regions.
\fi
}
\label{pix}
\end{center}
\end{figure}

Figure~\ref{pix} is a single event display which demonstrates the low
noise of the detector (average 6 hits for beam off events) 
and its clear multi-track capability. 
\iflindapicture
Six overlapping ring are shown.
\else 
This event has a $600\,\mbox{GeV/c}$ $\Sigma^-$ interaction, 
producing a $380\,\mbox{GeV/c}$ $\Lambda_c^+$, 
which decayed after $2.2\,\mbox{cm}$
into a $154\,\mbox{GeV/c}$ proton (track~4), a $152\,\mbox{GeV/c}$ $K^-$
(track~6), 
and a $74\,\mbox{GeV/c}$ $\pi^+$ (track~5). 
All three decay tracks are identified in the RICH detector.
\fi

A maximum likelihood analysis~\cite{likeli} is performed for each track 
in the event.  The 
algorithm uses tracking information to determine the ring
centers and then examines hypotheses for several different particle types
for each track. The likelihood function compares the number of seen hits 
to the expected number, using measured efficiencies ($N_0$), single
hit resolutions, and the track momentum, for every hypothesis. The expected
background is calculated for every track separately counting hits in a band
outside the radius for a \hbox{$\beta=1$} particle; this method is used because
it automatically takes into account overlapping rings.
To discriminate between different particles, a cut is used on the ratio of 
likelihoods for different hypotheses.
This method works for all hypotheses, even those below threshold.

\subsection{Determination of $N_0$}
\begin{figure}[htb]
\begin{center}
\leavevmode
\epsfxsize=\hsize
\epsfbox{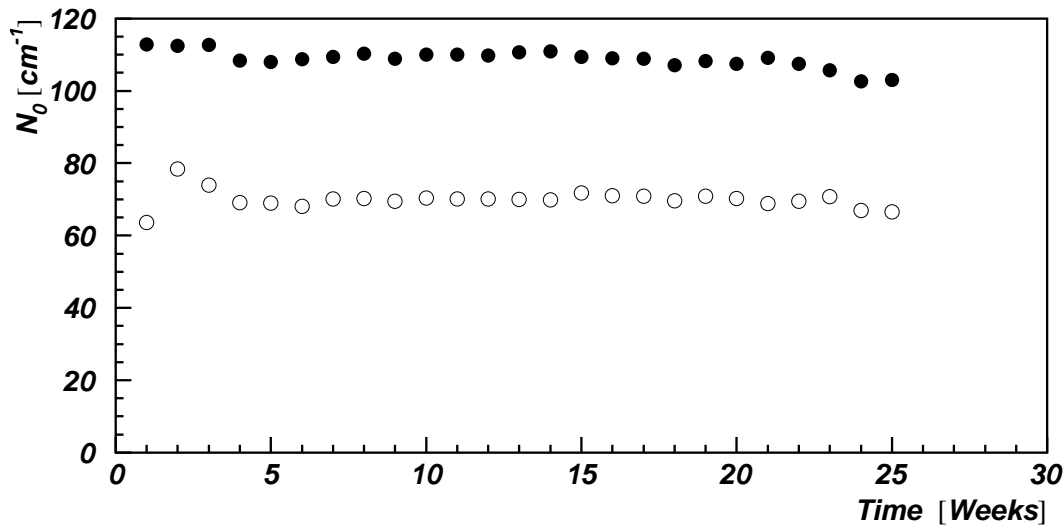}
\caption{$N_0$ for Center and West Region versus Time.}
\label{n0time}
\end{center}
\end{figure}

One measure of the response of a Cherenkov counter is the
Figure of Merit $N_{0}$, which is defined by the relation~\cite{desrich}
     $ N_{0} = N/(L \sin^{2} \theta)$ , where $N$ is the
number of detected photons, $L$ is
the radiator length and $\theta$ is the Cherenkov angle.
The photodetector has three separate regions: the central, high momentum
section and the two outer, low momentum sections, so three different
$N_{0}$'s must be calculated.  In addition, one of the central mirrors
has a section
removed where the non-interacting $600\,\mbox{GeV/c}$ beam tracks
passed, so another $N_{0}$ is needed to describe this region.
Because it is desirable to measure  $N_{0}$ throughout the run, standard 
interaction data
are used, with cuts applied to select tracks that are well-isolated
and have all of their photons in the region of interest.
Figure~\ref{n0time} shows 
$N_{0}$ vs time over the course of the run for two representative regions
of the detector: the central region (Average $N_0 = 108.3\,\mbox{cm}^{-1}$)
and one of the side regions (Average $N_0 = 69.7\,\mbox{cm}^{-1}$).
Apart from
the first three data sets, which had limited statistics,  $N_{0}$ is seen to
vary by less than $6\,\mbox{\%}$ over the course of the run.  
After correcting for the
number of dead and noisy channels observed, $N_{0}$ is 
constant to $2.5\,\mbox{\%}$ as a function of time.  
Table~\ref{n0tab} gives the average uncorrected
$N_{0}$ as a function of region.
\begin{table}[htb]
\center
\begin{tabular}{|l|c|} \hline
             Region  &        $N_{0}$ [$\mbox{cm}^{-1}$] \\ \hline
             Positive   &          $69.7$ \\
             Central    &         $108.3$   \\
             Hole       &         $ 75.6$   \\
             Negative   &         $ 71.4$   \\ \hline
\end{tabular}
\caption{$N_0$ for different detector regions.}
\label{n0tab}
\end{table}
Even though the Hole region has an $N_{0}$ which is similar to that of
the side regions, only a small number of tracks populate it so that the
average $N_{0}$ over all tracks in the Central region 
is $104\,\mbox{cm}^{-1}$.

\subsection{Determination of the refractive index}
\begin{figure}[htb]
\begin{center}
\leavevmode
\epsfxsize=\hsize
\epsfbox{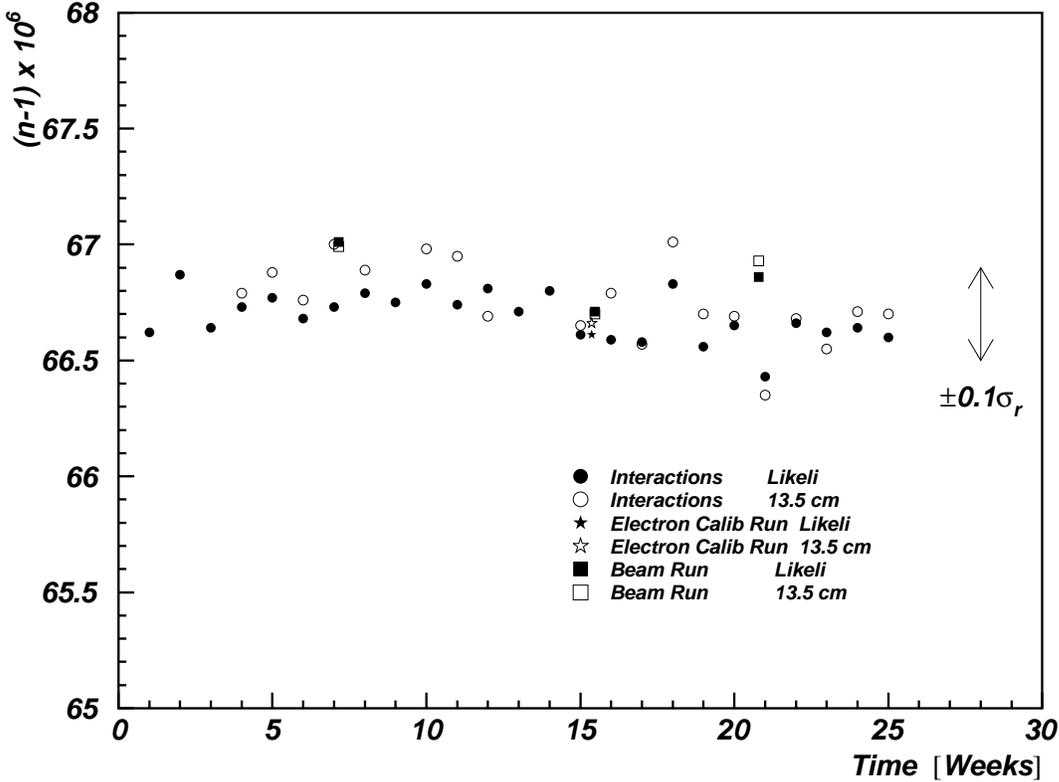}
\caption{$n-1$ versus Time, determined with different methods, which are
explained in the text. A $\pm 0.4\cdot 10^{-6}$ change in ($n-1$) 
changes the ring radius only by $10\,\mbox{\%}$ of the measured ring radius 
resolution $\sigma_r$.}
\label{nm1time}
\end{center}
\end{figure}

Determining the index of refraction of the gas $n$, or as done here ($n-1$), 
as a function of time  can be
done using standard interaction data, or special data sets such as 
electron beam data
and  $-600\,\mbox{GeV/c}$ $\pi^-$
beam data.  For interaction data, the tracks need to be isolated
so that one is not confused by nearby tracks or unassigned hits.  In all
data sets, in
order to avoid bias in the results from using the RICH determination of
the particle mass as much as possible, cuts from other detectors 
were applied.
For $-600\,\mbox{GeV/c}$ beam data, 
the Beam Transition Radiation Detector (BTRD) was used to select
$\pi^-$.  For the electron beam data, both the BTRD and another 
TRD (ETRD) were used
to select electrons.  For interaction data, the momentum was limited to
$100\,\mbox{GeV/c}$ and below and the ETRD was used to eliminate electrons.   
Kaons and heavier mass particles were eliminated if these hypotheses from
the likelihood analysis were selected as the most probable.  All other
tracks were kept and assigned the pion mass.
Data on ($n-1$) versus time from interaction data, the electron
beam data 
and a few randomly selected $-600\,\mbox{GeV/c}$ beam data is shown in
Fig.~\ref{nm1time}.  
Two different estimates of the ring radius
were used to calculate ($n-1$).  The first came from the Likelihood
determination, using all hits within a 
$3\,\sigma_h$ cut\footnote{$\sigma_h$ is the single hit resolution, 
determined in the next section.} of the most probable
radius.  The second used all hits within $13.5\,\mbox{cm}$ 
of the predicted track
center.  The second method does not rely on the likelihood analysis at
all, but obviously has much more background.  Shown on the figure is
the range of ($n-1$) needed to change the central
value of a $\beta=1$ ring radius distribution by 
$\pm 0.1\,\sigma_r$. 
Here the ring radius resolution $\sigma_r$ was taken to be $1.75\,\mbox{mm}$,
the average of the isolated single track resolution and the
value from the non-isolated track study, as described later.
As can be seen in the figure, the index of refraction of the gas was
essentially constant over the entire run.  

\subsection{Determination of the single hit resolution $\sigma_h$}
A basic parameter of the likelihood analysis is the single hit resolution.
This is determined by histogramming the radial distance of all hits from
a given track center which is determined by  projected tracks from an
upstream PWC system to the photocathode.
The width of the resulting 
distribution is then the single hit resolution $\sigma_h$. Care has to be taken
to select tracks that are of a similar kind, so
that the distribution is not broadened by track-to-track differences such
as momentum, particle type, etc.
The data set used for this analysis was taken to calibrate
another detector in the experiment (a lead glass detector), and consists
of a set of $60-80\,\mbox{GeV/c}$ single electron tracks (electron beam data)
which populate the horizontal
acceptance of the RICH.  Because the electrons were well identified in two
separate TRDs, there are no broadening
effects for these $\beta=1$~particles.  
The measured widths of the single hit radius distributions are summarized
in table~\ref{tab:snglhit}.
\begin{table}[htb]
\center
\begin{tabular}{|l|c|c|} \hline
\multicolumn{1}{|c|}{Track type}  & 
\multicolumn{2}{c|}{$\sigma_h$ [$\mbox{mm}$]} \\
  & Electron beam & Interaction \\ \hline
All tracks                      &     $5.40$   &     $5.75$            \\
All light on one mirror         &     $5.25$   &     $5.67$             \\
All light from multiple mirrors &     $5.64$   &     $5.77$              \\
\hline
\end{tabular}
\caption{RMS Widths of the single hit distance to ring center 
for different subclasses of tracks 
for electron beam and interaction data.}
\label{tab:snglhit}
\end{table}
The contribution to this
resolution from the finite precision of the mirror 
alignment procedure ($\sigma_m$) can
be estimated from the quadratic difference of the resolution of radial
distributions for tracks which had all of their light  on a single mirror
and for tracks which had their light distributed
over more than one mirror,
calculated from 
table~\ref{tab:snglhit} 
to be $\sigma_m = 2.06\,\mbox{mm}$.
The contribution to the single hit resolution from the finite precision
of the PWC tracking system ($\sigma_t$)
can be estimated by first fitting all of the
found hits to a circle and then using the center of this circle (and
not the predicted track center) to calculate the single hit resolution.
The quadratic difference of this resolution, measured to be $4.81\,\mbox{mm}$,
and that
found by using the projected track is then a measure of the tracking
system contribution, calculated to be $\sigma_t = 2.45\,\mbox{mm}$ 
(from chamber geometry, the expectation is $\sigma_t \approx 3.0\,\mbox{mm}$).

A second data set, taken from
interaction data, can also be used to measure the single hit resolution.
For this data set,
one needs to select a small momentum range in which the radii for
pion and kaon particles are well separated, here taken to be 
$60-65\,\mbox{GeV/c}$.
Electrons were removed from the sample with a TRD cut.
The resulting rms widths of the single hit radius distributions are shown in
table~\ref{tab:snglhit}.
The single hit resolution 
measured from this data sample is $\sigma_h=5.75\,\mbox{mm}$.
Because this data is restricted to a limited section of the phase space of
the detector,  for further analysis 
a single hit resolution from a
weighted average of the two data sets of
$\sigma_h=5.5\,\mbox{mm}\pm 0.1\,\mbox{mm}$ 
is used. This can be compared
to an estimate of this quantity calculated from the known physical
parameters of the detector, given in Table~\ref{snglhittab}.
\begin{table}[htb]
\center
\begin{tabular}{|l|c|} \hline
\multicolumn{1}{|c|}{Source}      &   Value [$\mbox{mm}$] \\ \hline
Pixel Size (Spacing/4)           &     $4.03$           \\
Mirror Alignment ($\sigma_m$)    &     $2.06$            \\
PWC Resolution ($\sigma_t$)      &     $3.0$             \\
Dispersion in Neon~\cite{bideau} &     $1.2$           \\ \hline
Total expected ($\sigma_h$)      &     $5.54$            \\ \hline
Total measured ($\sigma_h$)      &     $5.5\pm 0.1$     \\ \hline
\end{tabular}
\caption{Contributions to the single hit resolution $\sigma_h$.}
\label{snglhittab}
\end{table}

\subsection{Determination of the ring radius resolution $\sigma_r$}
Each phototube that fires gives an independent measurement of the ring
radius, when the ring center is given by the measured track parameters.
Thus the ring radius resolution $\sigma_r$ should be given by
\begin{equation}
   \sigma_r = \frac{\sigma_h}{\sqrt{N}} \label{eq:npmts}
\end{equation}
where  $\sigma_h$ is the single hit resolution determined above, and $N$ is 
the number of hits used for this ring. This can be
measured in the electron beam data previously discussed.
This is shown in 
table~\ref{tab:sngltrkelec}
which shows the rms widths for the ring radius distributions
and the number of hits found on these rings 
for both a simple radius fit
and for the Likelihood analysis output.
For these distributions the single hit
resolution of $\sigma_h=5.5\,\mbox{mm}$ was used and a 
$3\,\sigma_h$ cut was applied to remove close
noise hits.  Using equation~(\ref{eq:npmts}) above, one expects a ring
radius resolution of $\sigma_r\approx 1.5\,\mbox{mm}$, 
which is slightly larger than what is
observed.
\begin{table}[htb]
\center
\begin{tabular}{|l|c|c|c|} \hline
                          &          & & Expectation from \\
Method              & $N$    & $\sigma_r$ [$\mbox{mm}$] &
equation~(\ref{eq:npmts}) [$\mbox{mm}$]  \\ \hline
Radial Fit        & $13.29$  &  $1.46$  & $1.51\pm 0.03$          \\
Likelihood Fit    & $13.20$  &  $1.42$  & $1.51\pm 0.03$          \\ \hline
\end{tabular}
\caption{Number of hits and rms width of ring radius 
distributions for electron beam data.}
\label{tab:sngltrkelec}
\end{table}

\begin{figure}[htb]
\begin{center}
\leavevmode
\epsfxsize=\hsize
\epsfbox{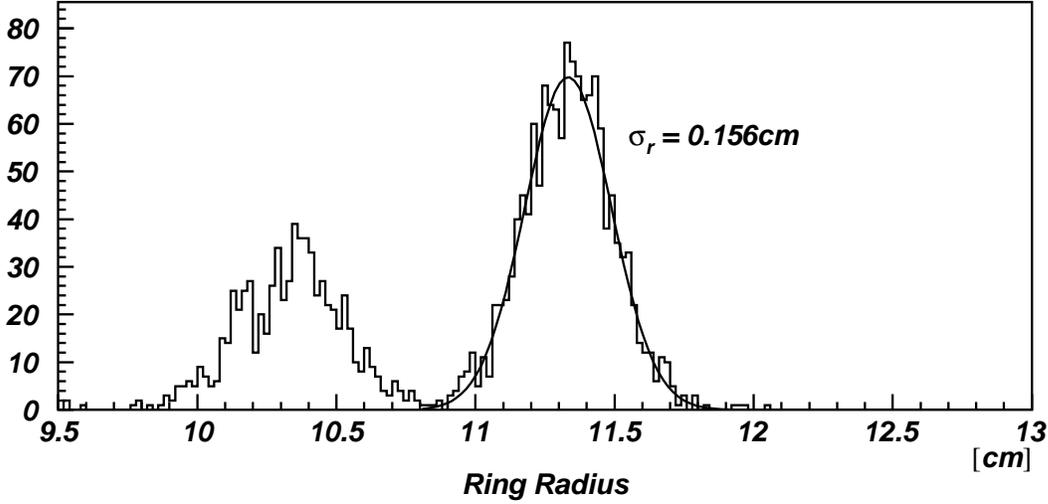}
\caption{Ring radius 
distributions for interaction data for tracks with $95-105\,\mbox{GeV/c}$
momentum. Very well separated peaks corresponding to pions (right) and
kaons (left) can be seen.}
\label{sngltrkinter}
\end{center}
\end{figure}

\looseness=+1
Measuring the single track ring radius resolution in this manner
in interaction data is difficult because of the complication of overlapping
tracks. Additional cuts have to be applied to isolate the tracks.  Two
cuts are applied: 1) the closest track must be at a distance of at
least two electron radii
away from the one considered and 2) the expected background 
from non-associated hits near the track
considered must be no larger than 1.  Again the single hit resolution was
set to $\sigma_h=5.5\,\mbox{mm}$ and a 
$3\,\sigma_h$ cut was used to constrain the number of hits
assigned to tracks. When
this is done, the resolution found is similar to that from single track
events.  
This is shown in Fig.~\ref{sngltrkinter}, which plots 
the ring radius distribution from $95-100\,\mbox{GeV/c}$ isolated tracks.
For these data, and for 
isolated $60-65\,\mbox{GeV/c}$ pion tracks, the results are summarized 
in table~\ref{tab:sngltrkinter}.
\begin{table}[htb]
\center
\begin{tabular}{|l|c|c|c|} \hline
                          &          & & Expectation from \\
Momentum Range            & $N$      & $\sigma_r$ [$\mbox{mm}$] & 
equation~(\ref{eq:npmts}) [$\mbox{mm}$]  \\ \hline
$95-105\,\mbox{GeV/c}$    & $13.18$  &  $1.56$  & $1.51\pm0.03$ \\
$60-65\,\mbox{GeV/c}$     & $11.22$  &  $1.74$  & $1.64\pm0.03$ \\ \hline
\end{tabular}
\caption{Number of hits and rms width of ring radius 
distributions for interaction data.}
\label{tab:sngltrkinter}
\end{table}
\begin{figure}[htb]
\begin{center}
\leavevmode
\epsfxsize=\hsize
\epsfbox{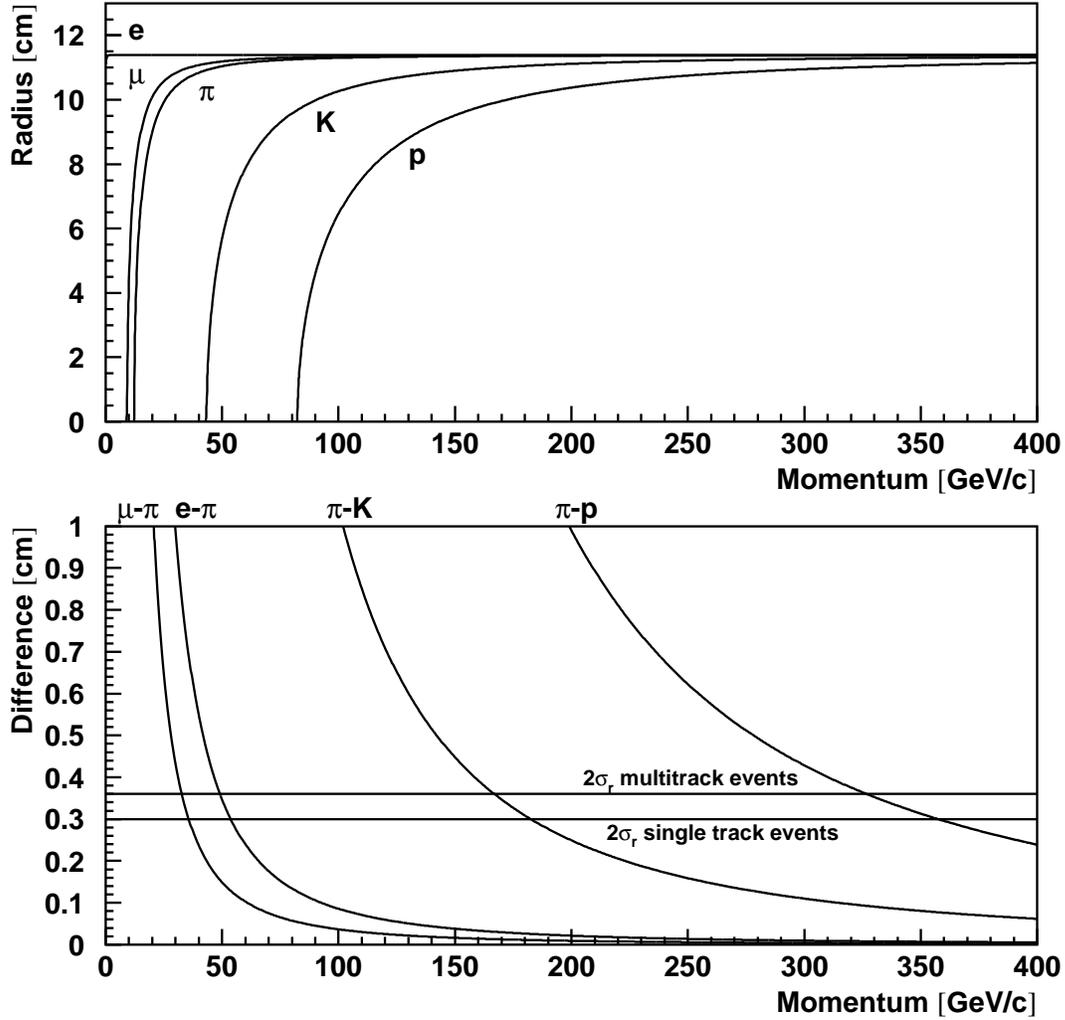}
\caption{Ring radii (top)  and separation (bottom) for different particles.
The two horizontal lines on the lower plot show the achieved
resolutions for single track and multitrack events, respectively.} 
\label{pid_fig_8}
\end{center}
\end{figure}
\noindent
Here, the expectations from equation~(\ref{eq:npmts}) are somewhat smaller
than what is observed.
Figure~\ref{pid_fig_8}
shows the expected ring radii at the photodetector
and the separation
between different ring radii for various particle
types as a function of momentum.
A single track resolution of $\sigma_r=1.5\,\mbox{mm}$ gives 
a $2\,\sigma_r$ $\pi$-$K$ separation out to about $180\,\mbox{GeV/c}$.

\section{Efficiencies for Particle Identification}
\looseness=1
To measure the efficiency for particle identification, a sample
of $\Lambda\rightarrow p\pi^-$ decays was used, 
which can be reconstructed without
using the information from the RICH detector, and determined how often the
RICH identifies the positive track as a proton.  This study was performed
in bins of $4\,\mbox{GeV/c}$ in the proton momentum, and a total of
37000~$\Lambda$ was used.  For  every momentum bin, the $\Lambda$ peak 
was fit before
and after a RICH cut\footnote{Likelihood of the track to be a 
proton has to be at least as big as to be a pion.}
was applied to the positive track.
\begin{figure}[htb]
\begin{center}
\leavevmode
\epsfxsize=\hsize
\epsfbox{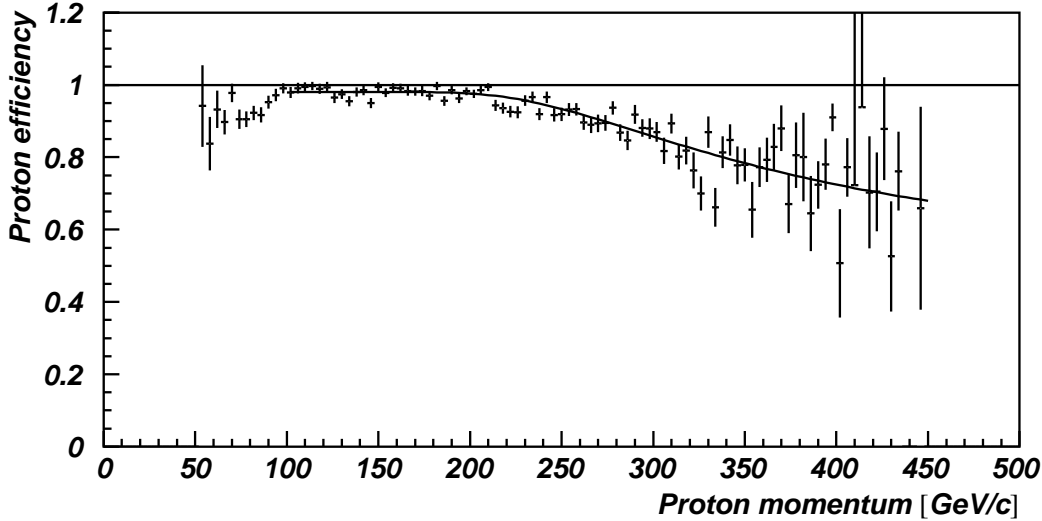}
\caption{Efficiency for identifying a proton as function of its momentum.
The likelihood of the track to be a proton has to be at least as big as to be
a pion.}
\label{proton_eff}
\end{center}
\end{figure}
In Fig.~\ref{proton_eff} it can be seen that the efficiency above the proton
threshold is 
well above $95\,\mbox{\%}$.
Close to and below the 
proton threshold ($\approx 90\,\mbox{GeV/c}$) the efficiency
is slightly lower; in this region, the likelihood algorithm is very sensitive
to fluctuations in the number of expected background hits.
At higher momenta the
proton and pion rings start to overlap.  A fit which takes these overlaps 
into account (line in Fig.~\ref{proton_eff})
gives a ring radius resolution of $\sigma_r=1.8\,\mbox{mm}$, which includes all
effects such as overlapping rings from different tracks,
tracking errors, etc, and because of these
effects the ring radius resolution for multi-track events is larger
than the ring radius resolution determined from isolated tracks.
With the observed resolution it is possible to 
separate (on a $2\,\sigma_r\,$level) 
kaons and pions up to $165\,\mbox{GeV/c}$ and protons and pions up to
$323\,\mbox{GeV/c}$. These numbers are expected to improve with ongoing
work on the track reconstruction code, which should reduce the contribution
of the PWC resolution (see table~\ref{snglhittab}) to a negligible level. 

\begin{figure}[htb]
\begin{center}
\leavevmode
\epsfxsize=\hsize
\epsfbox{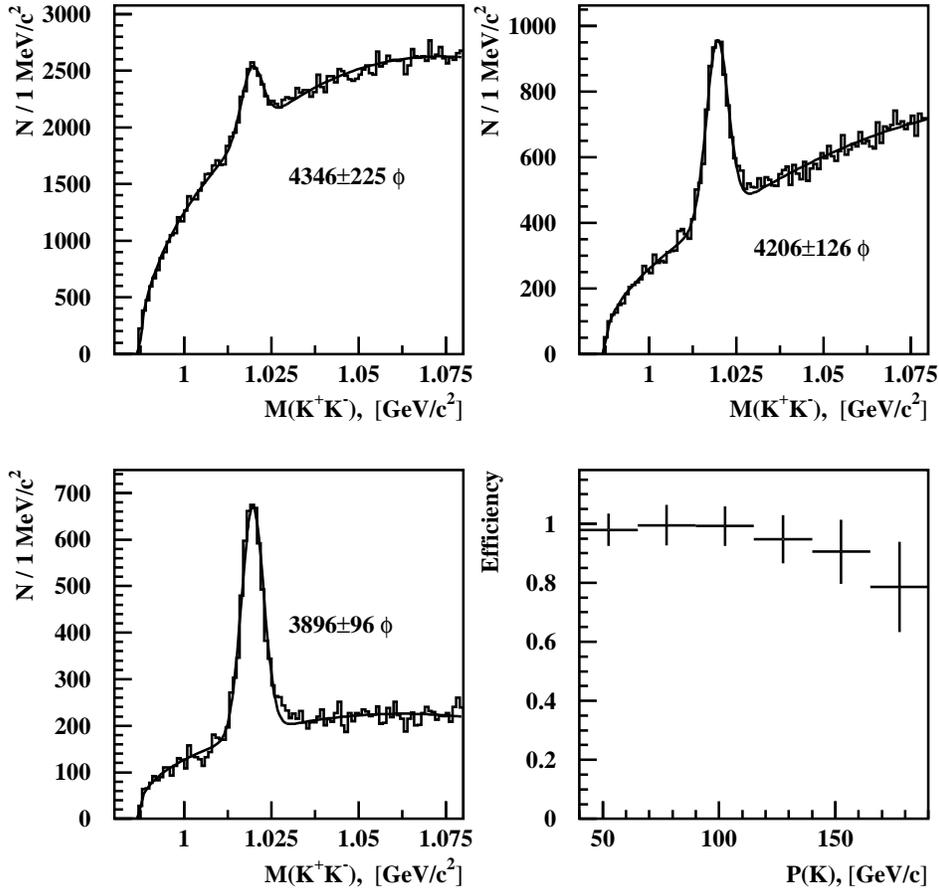}
\caption{Invariant mass of $K^+K^-$\@. Top left: No identification of
the kaons. Top right: One kaon is identified. Bottom left: Both kaons
identified. Bottom right: Efficiency for the RICH identifying a kaon
as a function of its momentum. For identification, the likelihood of the track
to be a kaon is required to be at least as big as to be a pion.}
\label{phi}
\end{center}
\end{figure}

\looseness=1
Figure~\ref{phi} shows the invariant mass for the two kaons in the
reaction $\Sigma^- +A\rightarrow A+\Sigma^- K^+K^-$.  The event has exactly 
two negative and one positive outgoing tracks, and the energy of the
outgoing particles is equal to the beam energy.
A clear peak of the $\Phi$ can be seen, and the RICH identification
has a very high efficiency for identifying the kaons. This confirms the 
proton result.

\looseness=1
To measure the mis-identification rate, the $\Lambda$ sample was used again,
asking how often the $\pi^-$ is identified as an antiproton or kaon, 
requiring that
the likelihood of the track to be a pion is smaller than the
maximum of the likelihoods to be a proton or kaon. 
\begin{figure}[htb]
\begin{center}
\leavevmode
\epsfxsize=\hsize
\epsfbox{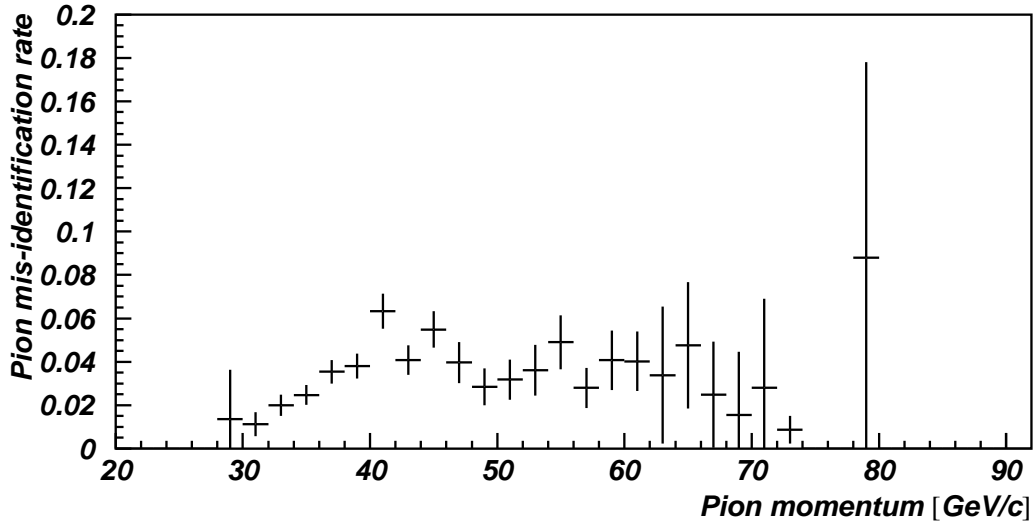}
\caption{Mis-identification rate (rate for identifying a pion as a 
proton or kaon) as a function of its momentum.
The likelihood of the track to be a pion has to be smaller than the
maximum of the likelihoods to be a proton or kaon.}
\label{pion_ineff}
\end{center}
\end{figure}
The result can be seen in 
Fig.~\ref{pion_ineff} to be only a few percent. 
For hypotheses below threshold the algorithm is 
sensitive to fluctuations in the background and to wrong ring center 
predictions by the track finding algorithm.

\begin{figure}[htb]
\begin{center}
\leavevmode
\epsfxsize=\hsize
\epsfbox{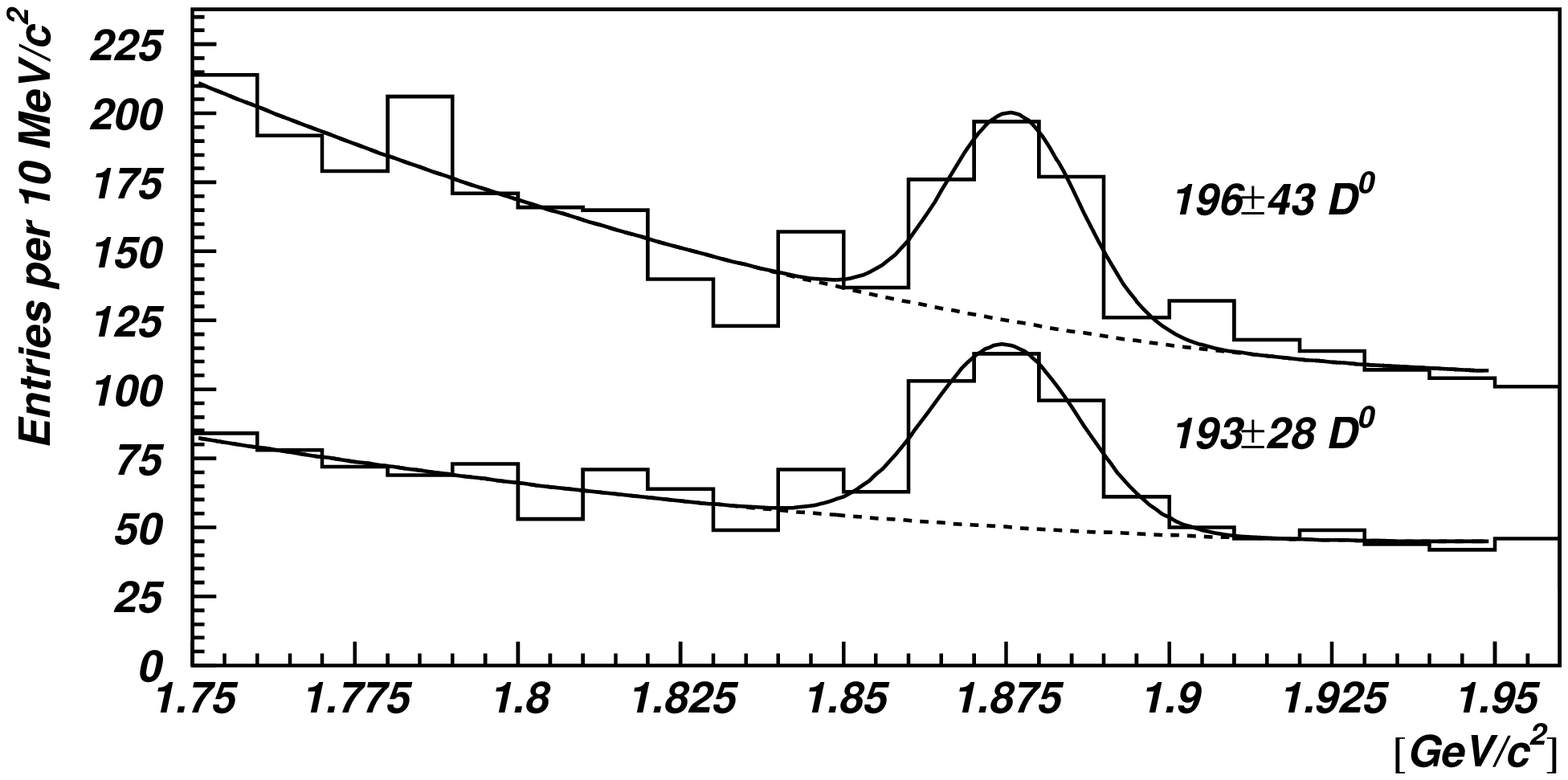}
\caption{Invariant mass of $K^- \pi^+$ (+cc) without and with RICH identified
kaon.}
\label{d0signal}
\end{center}
\end{figure}

In Fig.~\ref{d0signal} the use of the RICH detector in the
charm data analysis of SELEX is demonstrated.
The statistics available here is only a fraction of the experiment data
(due to the high efficiency of the RICH detector, a particle identification
cut was used during most of the first stage processing of the data).
A clear background suppression by a factor of 2.5 without a loss of signal 
(within the errors)  
is seen.  This confirms the previous results that the RICH detector is very
efficient. 

\clearpage
\section{Conclusions}
In this article we described the construction and performance of the SELEX
RICH detector. Stable performance of the detector was demonstrated over the
15~months of running time of the experiment. The refractive index did not
change at all, and the Figure of Merit $N_0$, which measures the efficiency of
detecting photons, only varied slightly.
It is on average for the central region
$104\,\mbox{cm}^{-1}$, equivalent to $13.6$~hits detected for a 
$\beta=1$ particle.

The measured single hit resolution of $\sigma_h=5.5\,\mbox{mm}$ can be 
explained totally by known factors and is dominated by the size of
the phototubes; with ongoing work on the track reconstruction code,
the second biggest contribution should be reduced significantly,
and $\sigma_h$ should approach $4.8\,\mbox{mm}$.

The ring radius resolution for isolated tracks can be explained totally
by equation~(\ref{eq:npmts}), and gives $\sigma_r=1.5\,\mbox{mm}$.  For 
interaction data, due to the overlapping of tracks, 
$\sigma_r=1.8\,\mbox{mm}$ was obtained, 
which corresponds to about $1.6\,\mbox{\%}$ ring radius resolution
at $\beta=1$.  This allows (on a $2\,\sigma_r\,$level) the separation of
kaons and pions up to $165\,\mbox{GeV/c}$ and of protons and pions up to
$323\,\mbox{GeV/c}$.

The efficiency of particle identification is very high 
($\approx 98\,\mbox{\%}$) and
only diminishes for high momenta as expected from the ring radius resolution.
Below threshold, the efficiency is well above $90\,\mbox{\%}$, with a
mis-identification rate of only a few percent.

All the performance parameters of the SELEX RICH are as good or even 
better than projected from simulation and prototype tests~\cite{protos}.
The results from this detector are used as one of the key tools in the 
analysis of the SELEX data.

\section{Acknowledgements}
This work could not have been performed without the help of a lot
of people.  We would like to thank the engineering and technical support staffs
at our institutes, especially at Fermilab, for their excellent work;
V.I.~Solyanik for his early contributions;
our summer students;
and the members of our collaboration 
and the Russian group leaders L.G.~Landsberg and E.M.~Leikin
for their continuous encouragement and support.

\end{document}